\documentclass[article,  twocolumn,showpacs,preprintnumbers,amsmath,amssymb]{revtex4}

\usepackage{graphicx}
\usepackage{dcolumn}
\usepackage{bm}

\begin{document}
\author{Xiaohua Wu} \author{Ke Xu}
\address{Department of Physics, Sichuan University, Chengdu 610064, China.}
\title{ Partial Standard Quantum Process Tomography}
\begin{abstract}
    With the general theory of standard quantum process tomography, we shall develop a scheme to decide
    an arbitrary matrix element of $\chi$, which is in the Choi matrix representation, in a scalable way.
\end{abstract}
 \maketitle The characterization of the evolution
of a quantum system is one of the main tasks to accomplish to
achieve quantum information processing. A general class of methods,
which have been developed in quantum information theory to
accomplish this task is known as quantum process tomography (QPT)-
for a review of quantum tomography, see Refs. [$1-3$]. Under general
assumptions, the evolution of quantum system can be represented by a
linear, completely-positive map which can be written as
\begin{equation}
\varepsilon(\rho)=\sum_{m}E^{m}\rho (E^{m})^{\dagger}
\end{equation}
with $\sum_{m}(E^{m})^{\dagger}E^{m}=\textbf{\texttt{I}}$. Using a
fixed set of operators $\{\tilde{E}_i\}$, which form a basis for the
set of operators on the state space, so that $E^m=\sum_i e^m_i
\tilde{E}_i$, one may introduce the $\chi$ matrix representation
with $\chi_{ij}=\sum_{m}e^m_i(e^m_j)^*$ for its matrix elements, and
rewrite  (1) in the way like
\begin{equation}
\varepsilon(\rho)=\sum_{ij}\chi_{ij}\tilde{E}_i(\rho)\tilde{E}_j^{\dagger}.
\end{equation}
The task of QPT can be viewed as to decide $\chi$ via different
protocols. Standard quantum process tomography (SQPT) was the first
method  proposed [1,4,5]. It involves preparing a set of input
$\{\rho_j\}$ and measuring its output $\varepsilon(\rho_j)$ via the
quantum state tomography. With the denotations that
$\varepsilon(\rho_j)=\sum_k\lambda_{jk}\rho_k$ and
$\tilde{E}_m\rho_j\tilde{E}_m^{\dagger}=\sum_{j}\beta^{mn}_{ij}\rho_j$,
where $\beta^{mn}_{ij}$ are complex numbers which can be determined
by standard algorithms from linear algebra given the
$\{\tilde{E}_m\}$ operators and $\{\rho_j\}$ operators [1], one may
get the relation
\begin{equation}
\sum_{mn}\beta^{mn}_{ij}\chi_{mn}=\lambda_{ij},
\end{equation}
where $\chi _{mn}$ can be determined by the given parameters
$\beta^{mn}_{ij}$ and the data of $\lambda_{ij}$ known from quantum
state tomography. Besides SQPT, other methods like the
ancilla-assisted quantum process tomography have also been proposed
[6].

Usually, the complete characterization of $\chi$ matrix is a
non-scalable task: For the  N d-levels system, there are about
$d^{4N}$ elements to be decided. Recently, a series of works have
demonstrated that it is possible to extract partial but nevertheless
relevant information about the quantum process in an efficient and
scalable way [7-13]. These approaches share an essential feature
that: They are based on the idea that the tomography of a quantum
map can be efficiently performed by studying certain properties of a
twirling of such a map. Another method, the so-called direct
characterization of quantum dynamics (DCQD), was also constructed
for use in partial characterization  of quantum dynamics [14-16].

Here, we shall focus on the  topic whether the general theorem of
SQPT developed in [1] can be applied for partial QPT. The case,
where the $\chi$ is defined by the Choi operators, should be
discussed in present work. With the general theory of SQPT in [1],
we shall develop a scheme to decide an arbitrary $\chi$ matrix
element, which certainly carries some partial information about the
quantum process [13], in a scalable way.  We shall show that: For
the $\chi$ in the Choi matrix representation, to decide its diagonal
element one need just a single measurement while to decide its
off-diagonal we require about sixteen measurements. This result is
independent of the actual dimension of the system. Our scheme also
represents a new kind of method of partial QPT without of  ancilla.
 In following argument, we shall
to show how to construct our scheme in detail.

\emph{The general theorem}.- With $\{\vert a\rangle\}_{a=0,1,...,
D-1}$ the basis of the D-dimensional system, we introduce the Choi
matrix
\begin{equation}
\tilde{E}_{ab}=\vert a\rangle\langle b\vert.
\end{equation}
This set of operators, $\{\tilde{E}_{ab}\}_{a,b=0,1,...,D-1}$, forms
an orthogonal basis with the algebra that
$\texttt{Tr}(\tilde{E}_{ab}^{\dagger}\tilde{E}_{cd})=\delta_{ac}\delta_{bd}$.
Any  $D\times D$ matrix, say, M, can be expanded with it,
$M=\sum_{ab=0}^{D-1}M_{ab}\tilde{E}_{ab}$, where the expanding
coefficients  $M_{ab}=\texttt{Tr}(\tilde{E}_{ab}^{\dagger}M)$ are
just the matrix elements of M. Let each Kraus-operator in $\{E^m\}$
to be expanded with the conjugated Choi matrices in (6),
$E^m=\sum_{e,f=0}^{D-1}e^m_{ef}{\tilde{E}}_{ef}$ with
$e^m_{ef}=\texttt{Tr}[\tilde{E}_{ef}^{\dagger}E^m]$ and
$(E^m)^{\dagger}=\sum_{g,h=0}^{D-1}(e^m_{gh})^*(\tilde{E}_{gh})^{\dagger}$,
we can define the $\chi$ in the Choi matrix representation with its
elements to be
\begin{equation}
\chi_{ef;gh}=\sum_m e^m_{ef} (e^m_{gh})^*.
\end{equation}
We call $\chi_{ef;gh}$ the diagonal matrix element if $e=g$ and
$f=h$, else, we call it the off-diagonal matrix element.

  The relation in
(3), which is one of the main results of the general theorem of SQPT
[1], puts no constraints on our choices to define $\lambda$ and
$\chi$ matrices. In present work, we use $\{\tilde{E}_{ab}\}$ to
describe the input and $\varepsilon(\tilde{E}_{ab})$ for its output.
Let $\varepsilon(\tilde{E}_{ab})$ be expanded with the  set of Choi
operators in (4),
$\varepsilon(\tilde{E}_{ab})=\sum_{c,d=0}^{D-1}\lambda_{ab;cd}\tilde{E}_{cd}$
with its elements  to be
\begin{equation}
\lambda_{ab;cd}=\texttt{Tr}[\tilde{E}_{cd}^{\dagger}\varepsilon(\tilde{E}_{ab})].
\end{equation}
 With the
above definitions in hands, it can be shown that there exists a
one-to-one mapping between the $\lambda$ matrix element and the
$\chi$ matrix element,
\begin{equation}
\lambda_{ab;cd}=\chi_{ca;db}.
\end{equation}
This result can be proved by the general theory of SQPT. From the
expressions of $\lambda_{ab;cd}$  and  $\chi_{ef;gh}$ in above,
there should be
\begin{equation}
\lambda_{ab;cd}=\sum_{e,f,g,h=0}^{D-1}\beta^{ef;gh}_{ab;cd}\chi_{ef;gh}
\end{equation}
where the  transformation matrix $\beta$ has its elements to be
\begin{equation}
\beta^{ef;gh}_{ab;cd}=\texttt{Tr}(\tilde{E}_{cd}^{\dagger}\tilde{E}_{ef}\tilde{E}_{ab}\tilde{E}_{gh}^{\dagger}).
\end{equation}
 Each matrix
element of $\beta$ can be directly calculated,
$\beta^{ef;gh}_{ab;cd}=\texttt{Tr}[\vert d\rangle\langle
c\vert)\cdot\vert e\rangle\langle f\vert\cdot \vert a\rangle\langle
b\vert\cdot(\vert h\rangle\langle
g\vert)]=\delta_{ec}\delta_{fa}\delta_{gd}\delta_{hb}$. With columns
indexed by $ef;gh$ and rows by $ab;cd$, $\beta$ can be expressed as
a $D^4\times D^4$ matrix. In each row and each column of it, there
is just one non-zero matrix elements (with the value of 1).
Furthermore, one may verify that the determinant of $\beta$ equals 1
while its inverse is the transpose of itself,
\begin{equation}
\det(\beta)=1, \beta^{-1}=\beta^{\texttt{T}}.
\end{equation}
 According to the famous
Cramer's rule [17]: \emph{If
$\texttt{\textbf{A}}\texttt{\textbf{x}}=\texttt{\textbf{b}}$ is a
system of n linear equations in n unknowns such that $\det
(A)\neq0$, then the system has a unique solutions. } Equation (8)
can be viewed as such a set of linear equations with:
$\texttt{\textbf{A}}\rightarrow \beta$,
$\texttt{\textbf{x}}\rightarrow \chi$, and
$\texttt{\textbf{b}}\rightarrow \lambda$. From (10), we conclude
that $\chi$ is unambiguously determined by $\lambda$,
$\chi=\beta^{\texttt{T}}\lambda$, since $\det(\beta)=1$. Finally,
using the known result of $\beta^{ef;gh}_{ab;cd}$ for (8), we shall
get the one-to-one mapping described by (7).

As a direct application of the one-to-one mapping in (8), we find
that the diagonal matrix element  $\chi_{ab;ab}(\lambda_{bb;aa})$ is
physical meaningful: it represents the probability of the transition
from the initial a-th level to the final b-th level,
\begin{equation}
\chi_{ab;ab}=\langle b\vert\varepsilon(\vert a\rangle\langle
a\vert)\vert b\rangle.
\end{equation}
In other words, the classical measurement of the transition
probability of the D-levels atom can be also viewed as to decide the
diagonal $\chi$ matrix element in the scheme of SQPT. With a simple
reasoning, we have
\begin{equation}
\texttt{Tr}\chi=\sum_{a,b=0}^{D-1}\chi_{ab;ab}=\sum_{a=0}^{D-1}\texttt{Tr}[\varepsilon(\vert
a\rangle\langle a\vert)].
\end{equation}
For the trace-preserving cases, $\texttt{Tr}\chi=D$.

\emph{Choi matrix  SQPT.}-As we have shown, the $\chi$ matrix in (5)
should be given if the $\lambda$ matrix in (6) has been decided.
However, the $\lambda$ matrix elements,
$\lambda_{ab;cd}=\texttt{Tr}[\tilde{E}_{cd}^{\dagger}\varepsilon(\tilde{E}_{ab})]$,
can not be directly measured if one of the operators,
$\tilde{E}_{cd}^{\dagger}$ and $\tilde{E}_{ab}$, is non-Hermitian.
This problem can be solved with the following protocol: At first,
one may introduce a set of linearly independent states,
$\{\vert\Psi_m\rangle\}_{m=1,...,D^2}$, and expand each
$\tilde{E}_{ab}$ as
\begin{equation}
\tilde{E}_{ab}=\sum_{m=1}^{D^2}r^{ab}_m\vert\Psi_m\rangle\langle\Psi_m\vert
\end{equation}
where the coefficients $r^{ab}_m$ are known. Now,
$\varepsilon(\tilde{E}_{ab})=\sum_{m=1}^{D^2}r^{ab}_m\varepsilon(\vert\Psi_m\rangle
\langle\Psi_m\vert)$ since that the operation of $\varepsilon$ is
linear. Then, giving  a  set of linear independent Hermitian
operators $\{\hat{O}_n\}_{n=1,...,D^2}$
($\hat{O}_n=\hat{O}_{n}^{\dagger}$), we rewrite
$\tilde{E}^{\dagger}_{cd}$ as
\begin{equation}
\tilde{E}^{\dagger}_{cd}=\sum_{n=1}^{D^2}s^{cd}_n\hat{O}_n
\end{equation}
with known parameters $s^{cd}_n$. Because that performing trace is
also a linear operation, there should be
$\texttt{Tr}[\tilde{E}_{cd}(...)]=\sum_{n=1}^{D^2}s^{cd}_n
\texttt{Tr}[\hat{O}_n(...)]$. Finally, the way of measuring
$\lambda_{ab;cd}(\chi_{ca;db})$ is clear,
\begin{equation}
\lambda_{ab;cd}=\sum_{m,n=1}^{D^2}r^{ab}_ms^{cd}_n\texttt{Tr}[\hat{O}_n\varepsilon(\vert\Psi_m\rangle
\langle\Psi_m\vert)].
\end{equation}
The configuration space of the measurements  is known,\[
\mathcal{M}:\{\texttt{Tr}[\hat{O}_n\varepsilon(\vert\psi_m\rangle\langle\psi_m\vert)]\}_{m,n=1,
2,...,D^2}.\] To perform the complete SQPT, we shall carry out all
$D^2\times D^2$ measurements in $\mathcal{M}$ and decide each
$\lambda(\chi)$ matrix element one by one according to their known
coefficients, $r^{ab}_i$ and $s^{cd}_j$, here. However, if only a
selected $\lambda_{ab;cd}(\chi_{ca;db})$ is to be measured, one may
just perform the measurements  satisfying the constraint that
$r^{ab}_n\cdot s^{cd}_m\ne0$.

In fact, it has been shown that each $\tilde{E}_{ab}$ can always be
expanded with four pure states [1]. Introducing
\begin{equation}
\vert ab,+\rangle=\frac{\sqrt{2}}{2}(\vert a\rangle+\vert b\rangle),
\vert ab,-\rangle=\frac{\sqrt{2}}{2}(\vert a\rangle+i\vert
b\rangle),
\end{equation}
for $a<b$,  each non-Hermitian $\tilde{E}_{ab}$ can  be expanded
with
\begin{eqnarray}
\tilde{E}_{ab}&=&\vert ab, +\rangle\langle ab,+\vert+i\vert
ab,-\rangle\langle ab, -\vert \nonumber\\
&&-\frac{1+i}{2}(\vert a\rangle\langle a\vert+\vert b\rangle\langle
b\vert).
\end{eqnarray}
For the case $a>b$, $\tilde{E}_{ab}$ can be directly derived with
$\tilde{E}_{ab}=\tilde{E}_{ba}^{\dagger}$. Formally, for $a\ne b$,
we have
$\tilde{E}_{ab}=\sum_{i=1}^4r^{ab}_i\vert\psi^{ab}_i\rangle\langle\psi^{ab}_j\vert$
 with $\forall\vert\psi_i^{ab}\rangle\in\{\vert a\rangle, \vert b\rangle, \vert ab,\pm\rangle\}$.
In the similar way,
 we expand $\tilde{E}_{cd}^{\dagger}$ with
$\tilde{E}_{cd}^{\dagger}=\sum_{j=1}^4s^{cd}_j\vert\psi^{cd}_j\rangle\langle\psi^{cd}_j\vert$
by requiring $\forall\vert\psi^{cd}_j\rangle\in\{\vert
c\rangle,\vert d\rangle,\vert cd,\pm\rangle\}$. Now, equation (15)
is simplified into
\begin{equation}
\lambda_{ab;cd}=\sum_{i,j=1}^4
r^{ab}_is^{cd}_j\langle\psi^{cd}_j\vert\varepsilon(\vert\psi_i^{ab}\rangle\langle\psi^{ab}_i\vert)\vert\psi^{cd}_j\rangle.
\end{equation}
To decide the off-diagonal matrix element of $\chi$,  we need
sixteen measurements by taking $\vert\psi^{ab}_i\rangle$ for the
input state and measuring its output state
$\varepsilon(\vert\psi_i^{ab}\rangle\langle \psi^{ab}_i\vert)$ with
the projective operator
$\vert\psi^{cd}_j\rangle\langle\psi^{cd}_j\vert$.

 It can be seen that the number of
all the possible input states, $\vert\psi^{ab}_j\rangle$ with $0\le
a< b\le D-1$ and j=1, 2, 3, 4, is limited to be $D^2$: The basis
vector $\vert a\rangle$ has D terms,
 while $\frac{\sqrt{2}}{2}(\vert a\rangle+\vert b\rangle)$ and $\frac{\sqrt{2}}{2}(\vert a\rangle+i\vert b\rangle)$
 has the same number of $\frac{1}{2}D(D-1)$, respectively. For convenience, we use
 $\mathcal{S}$ to denote all these states,
\[
 \mathcal{S}:\{\vert a\rangle,\frac{\sqrt{2}}{2}(\vert a\rangle +\vert b\rangle),
 \frac{\sqrt{2}}{2}(\vert a\rangle +i\vert b\rangle)\}_{0\le
 a<b\le D-1}.\] For the trace-preserving cases, only
$D^2(D^2-1)$ measurements  are independent. An interpretation for it
is like this: the following D measurements, $\langle
a\vert\varepsilon(\vert\psi^{ab}_i\rangle\langle\psi^{ab}_i\vert)\vert
a\rangle $ with a=0, 1,..., D-1, should be performed for a given
input state $\vert\psi^{ab}_i\rangle$. With $\sum_{a=0}^{D-1}\vert
a\rangle\langle a\vert=\texttt{\textbf{I}}_D$, we can always leave
 $\langle
D-1\vert\varepsilon(\vert\psi^{ab}_i\rangle\langle\psi^{ab}_i\vert)\vert
D-1\rangle$ unmeasured and decide its value according to\[ \langle
D-1\vert\varepsilon(\vert\psi^{ab}_i\rangle\langle\psi^{ab}_i\vert)\vert
D-1\rangle=1-\sum_{a=0}^{D-2}\langle
a\vert\varepsilon(\vert\psi^{ab}_i\rangle\langle\psi^{ab}_i\vert)\vert
a\rangle.\] This fact is consentient with our general analysis about
the $\chi$ matrix: For the D-dimensional system, it has $D^2(D^2-1)$
independent real parameters and $D^2$ additional constraints for
trace.

 \emph{The N d-levels system}.- The way of performing partial QPT is
 important for the
 the N d-levels system  specially when d is a large number. For such cases, to performing the complete
 QPT becomes a non-scalable task since the
 number of the required experiments is exponentially increased with
 N. Our general protocol also holds for  the N d-levels system by taking it as
 a special case of $D=d^N$. With
 \begin{equation}
 \vert a\rangle=\vert a_1\rangle \otimes\cdot\cdot\cdot\otimes\vert
 a_j\rangle\otimes\cdot\cdot\cdot\otimes\vert a_N\rangle
\end{equation}
where $\{\vert a_j\rangle \}_{a_j=0,1,...d-1}$ is the basis of the
j-th d-dimensional subsystem, we could define a relation between the
single index with its corresponding  string of local indices, say,
$a\rightarrow a_1a_2\cdot\cdot\cdot a_N$, $b\rightarrow
b_1b_2\cdot\cdot\cdot b_N$, \emph{etc..} If the  set of product
basis $\{\prod_{j=1}^N\otimes\vert a_j\rangle\}$ is to be used
instead of $\{\vert a\rangle\}$ used above, all the equations in the
product basis can be easily given by substituting each single index
with its corresponding string of local indices. For example,
 the  one-to-one mapping  in (7)  with the local indices should be,
\begin{equation}
\lambda_{a_1\cdot\cdot\cdot a_N b_1\cdot\cdot\cdot b_N;
c_1\cdot\cdot\cdot c_N d_1\cdot\cdot\cdot d_N}
=\chi_{c_1\cdot\cdot\cdot c_N a_1\cdot\cdot\cdot
 a_N;d_1\cdot\cdot\cdot d_N b_1\cdot\cdot\cdot b_N}.\nonumber
\end{equation}

For the two qubits case, with $\vert
+\rangle=\frac{\sqrt{2}}{2}(\vert 0\rangle+\vert 1\rangle)$ and
$\vert -\rangle=\frac{\sqrt{2}}{2}(\vert 0\rangle+i\vert 1\rangle)$,
$\mathcal{S}$ contains following product states, $\vert
0\rangle\vert 0\rangle, \vert 0\rangle \vert 1\rangle, \vert 1
\rangle\vert 0\rangle, \vert 1\rangle \vert 1\rangle, $$\vert
0\rangle\vert \pm\rangle, \vert 1\rangle\vert \pm\rangle, \vert
\pm\rangle \vert 0\rangle, \vert \pm\rangle \vert 1\rangle,$ and the
four Bell-type states, $\frac{\sqrt{2}}{2}(\vert 00\rangle+\vert
11\rangle)$, $\frac{\sqrt{2}}{2}(\vert 00\rangle+i\vert 11\rangle)$,
$\frac{\sqrt{2}}{2}(\vert 01\rangle+\vert 10\rangle)$, and
$\frac{\sqrt{2}}{2}(\vert 01\rangle+i\vert 10\rangle)$. For the more
general N d-levels case with the  product basis, $\mathcal {S}$
contains a series of product states and M-parties maximally
entangled states with M arranged from 2 to N.
 Suppose we
have the freedoms of choosing  one of the followings two mappings,
$\vert a_i\rangle \rightarrow \vert\uparrow_i\rangle $, $\vert
b_i\rangle \rightarrow \vert\downarrow_i\rangle $ or $\vert
a_i\rangle \rightarrow \vert\uparrow_i\rangle $, $\vert b_i\rangle
\rightarrow \vert\downarrow_i\rangle $, for each site, by
renumbering the sequence of all the  N sites, we can  write $\vert
ab,\pm\rangle $ in (16) with
\begin{equation}
\vert ab,+\rangle=\vert
G^M\rangle\prod_{j=M+1}^N\otimes\vert\uparrow_j\rangle, \vert
ab,-\rangle=\vert
\tilde{G}^M\rangle\prod_{j=M+1}^N\otimes\vert\uparrow_j\rangle\nonumber
\end{equation}
where the  two Greenberger-Horne-Zeilinger type states, $\vert
G^M\rangle=\frac{\sqrt{2}}{2}(\vert\uparrow\rangle^{\otimes
M}+\vert\downarrow\rangle^{\otimes M})$ and $\vert
\tilde{G}^M\rangle=\frac{\sqrt{2}}{2}(\vert\uparrow\rangle^{\otimes
M}+i\vert\downarrow\rangle^{\otimes M})$, are the maximally
entangled states among M-parties.

 \emph{Discussion}-With the general theory of SQPT in [1], we have developed
 a scheme of performing partial SQPT  for $\chi$ in the
 Choi matrix representation. Besides the fact that the set of Choi
 operators is a convenient basis for an arbitrary D-dimensional
 system, $\chi$ matrix in this representation  is shown to be
 physical meaningful in the sense that its diagonal matrix element is just the transition probability
 of the system. For the N qubits case, it can be shown that
 $\chi^C$ (in the Choi matrix representation ) and $\chi^P$ (in the Pauli matrix representation), are equivalent
 with each other. An interpretation is like this: Let's at first
 consider the single qubit case, it can be seen that the set of Pauli operators is
 related with the Choi operators by a unitary transformation U,
\begin{equation}
\left(
  \begin{array}{c}
    \texttt{\textbf{I}} \\
    \sigma_x \\
    \sigma_y \\
   \sigma_z \\
  \end{array}
\right)=\left(
          \begin{array}{cccc}
            \frac{\sqrt{2}}{2} & 0 & 0 & \frac{\sqrt{2}}{2}\\
            0& \frac{\sqrt{2}}{2} & \frac{\sqrt{2}}{2} & 0 \\
            0 & -i\frac{\sqrt{2}}{2} & i\frac{\sqrt{2}}{2}& 0 \\
            \frac{\sqrt{2}}{2} & 0 & 0 & -\frac{\sqrt{2}}{2}\\
          \end{array}
        \right)\left(
                 \begin{array}{c}
                   \bar{E}_{00} \\
                   \bar{E}_{01} \\
                   \bar{E}_{10} \\
                   \bar{E}_{11} \\
                 \end{array}
               \right)\nonumber,
\end{equation}
where we define  $\bar{E}_{ab}=\sqrt{2}\tilde{E}_{ab}$. One may
easily verified that $UU^{\dagger}=\texttt{\textbf{I}}_4$ and get
the relation $\chi^P=U\chi^CU^{\dagger}$. For the N qubits case, if
$\chi^P$ is  expanded with the set of Pauli operators
$\{\prod_{k=1}^N\otimes \sigma^k\}$ while $\chi^C$ is defined with
$\{\prod_{k=1}^N\otimes \bar{E}^k\}$, the relation
$\chi^P=\tilde{U}\chi^C\tilde{U}^{\dagger}$ still holds with
$\tilde{U}=\prod_{k=1}^NU^k$. From above discussion, it's hard for
us to claim that one representation has privilege over the other
representation for defining  $\chi$ matrix.

From equation (15), it should be noted that the ways of performing
 complete SQPT are not limited, any set of linearly
independent  states $\{\vert\Psi_m\rangle\}$ with another set of
linearly independent Hermitian operators $\{\hat{O}_n\}$ can be
applied for this task. Among all the possible ways, our protocol in
(18) works for deciding an arbitrary $\chi$ matrix element with a
fixed number of measurements. If the complete SQPT should be
performed, one may use the following scheme: Let
$\{\vert\phi^k_j\}_{j=1,...,d^2}$ the set of linearly independent
states for the the $k-th$ subsystem, the input states $\vert
\Psi_m\rangle$ in (13) may be defined as a pure product states, say,
$\vert\Psi_m\rangle=\prod_{k=1}^N \otimes\vert\phi^k\rangle$;
Considering that $\hat{\Gamma}^k_0=\texttt{\textbf{I}}_d$ and
$\{\hat{\Gamma}^k_j\}_{j=1,...,d^2-1}$, where $\hat{\Gamma}^k_j$ is
the generator of $SU(d)$, forms a orthogonal basis of the k-th
subspace, the operators $\hat{O}_n$ in (14) can be chosen to be
$\hat{O}_n=\prod_{k=1}^N\otimes\hat{\Gamma}^k$, the complete SQPT
performed in this way has the unique property that it does not
require any two-bodies or multi-bodies interaction.

There are two related problems still unsolved with the present work.
At first, if all the experiment data are  used to reconstruct the
whole $\chi$ matrix, how to keep the positivity of $\chi$ is not
given since that the partial SQPT is concerned here. Second, our
protocol of partial SQPT is limited to the case where $\chi$ is in
the Choi matrix representation, it is still an open problem whether
the general theory of SQPT in [1] can  be also applied for the
partial QPT when $\chi$ is in other physical representations.

 Finally, let's make a short summary for our work.
 With the general theorem of SQPT, we developed a method to
estimate an arbitrary $\chi$ matrix element in the Choi matrix
representation.  Our  observation is that: To decide the diagonal
matrix element, one just needs a single  measurement; To decide an
arbitrary off-diagonal matrix element, we should carry out sixteen
measurements. This result is independent of the actual dimension of
the system. Compared with the known methods of partial QPT, our
scheme does not require any additional resources. It can be applied
for the case where a clean ancilla system is not available.

\end{document}